\documentclass{article}
\usepackage{spconf,amsmath,graphicx}
\usepackage[T1]{fontenc}
\usepackage{mathptmx}
\usepackage{booktabs}
\usepackage[hidelinks]{hyperref}
\usepackage{xurl}\Urlmuskip=0mu
\usepackage{amssymb}

\title{Simulated transcranial ultrasound of human language cortex: verifying heard speech needs far less data and signal than decoding it}
\name{Omer Shalev}
\address{Independent researcher}

\begin{document}
\maketitle

\begin{abstract}
Functional ultrasound could image cortical blood volume through the temporal bone, but the skull leaves a weak signal, and task-evoked cortical imaging through the intact adult skull has not yet been shown. We built a voxel-specific Monte Carlo link-budget simulation of a probe on the left temple and used it to degrade public fMRI of three people who heard 15 hours of stories. On the same anatomical voxels we trained a published state-of-the-art fMRI language decoder (Tang et al., 2023) and two verifiers that only check a candidate text. In clean data the verifiers needed far less training (11 to 12 minutes against 3.2 to 6.6 hours), but all three learned at a similar rate. Noise changed this qualitatively: the verifiers kept gaining evidence while the decoder slowed and, under simulated transcranial physics, stalled, whereas the verifiers still reached their goal under a favorable simulated link budget. This asymmetry suggests that low-SNR modalities such as transcranial ultrasound may support verification well before decoding.
\end{abstract}


\section{Introduction}
\label{sec:intro}

Functional ultrasound images cerebral blood volume with power Doppler \cite{mace2011}. In monkeys it has read out movement plans from single trials \cite{norman2021,griggs2024}, and in people it has recorded task responses through an acoustically transparent skull window \cite{rabut2024}. Through the intact adult skull, functional transcranial Doppler has long measured task-evoked blood-flow velocity in the large arteries, for example to lateralize language \cite{knecht1998}, but it does not image cortex. To our knowledge, contrast-free functional ultrasound imaging through the intact adult skull has been reported only for a breath-hold response in arteries around the midbrain of 13 adults \cite{vienneau2026}, and the same group measured coded-excitation gains of 11 to 18\,dB through the temporal window \cite{vienneau2023}. No task-evoked cortical response has yet been imaged through the temporal bone.

Here we asked what such a signal could support for language cortex listening to speech, and whether the answer depends on the readout. A decoder must produce the heard text from an open set; a verifier only scores a candidate text supplied from outside. fMRI and functional ultrasound both read the hemodynamic response, through blood oxygenation and blood volume, so fMRI degraded by ultrasound-sized noise can serve as a first approximation to a transcranial recording of the same cortex. Our contribution is a robustness map of neural language readouts under a simulated transcranial link budget. On the same anatomically defined voxels, and in one unit, bits of evidence per 20-s window, we trained a published decoder \cite{tang2023} and two verifiers, encode-then-compare and contrastive, on 11 minutes to 15 hours of data from three participants, with injected noise ranging from clean fMRI to the median and optimistic cases of a per-voxel Monte Carlo over six uncertain inputs. Three results follow. The verifiers need 16 to over 30 times less training data than the decoder. With 15\,h the decoder keeps its goal only to $-0.8$ to $-2.3$\,dB, encode-then-compare to $-14.5$ to $-16.5$\,dB and the contrastive verifier to $-17.6$\,dB or beyond. Noise changes the readouts qualitatively: in clean data all three learn at a similar rate, 2.6 to 4.2 bits per tenfold more data, but with noise the verifiers keep gaining evidence while the decoder slows (at $-6$\,dB it gains 1.0 to 1.8 bits per tenfold less, the verifiers at most 0.4 less) and, under simulated transcranial physics, stalls.

\begin{figure}[t]
\centering
\includegraphics[width=\columnwidth]{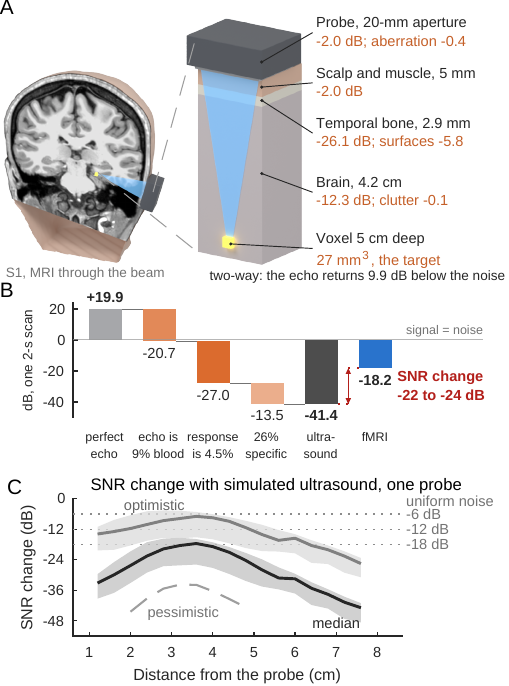}
\caption{\textbf{Through the temporal bone, simulated ultrasound leaves voxels far below fMRI's SNR.} (A) S1's MRI cut through the beam from the left temple to a voxel 5\,cm deep (seen from the front), and the path enlarged; two-way losses at 2\,MHz, Monte Carlo inputs at medians. (B) Its speech signal in one 2-s scan, step by step (rounded). (C) SNR change, $-10\log_{10}(1+f_v)$ (Eq.~\ref{eq:fv}), of the readouts' voxels with one probe: median and 10th to 90th percentile per 4-mm bin, mean of participants; dashed, pessimistic; dotted, the uniform noise of Fig.~\ref{fig:learning}.}
\label{fig:imaging}
\end{figure}

\begin{figure*}[t]
\centering
\includegraphics[width=\textwidth]{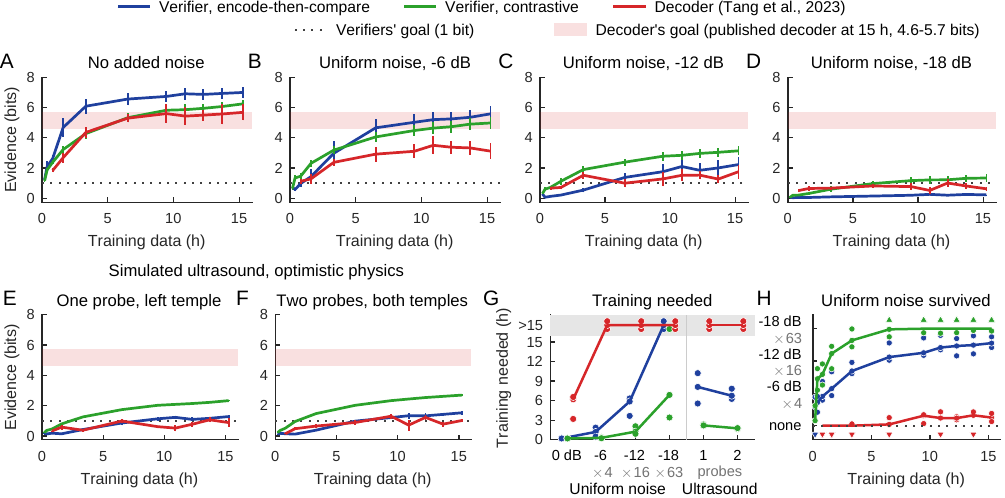}
\caption{\textbf{Verifiers need far less data, and keep learning under noise that stops the decoder.} (A to F) Bits of evidence per held-out 20-s window as training data grow; mean $\pm$ s.e.m., three participants. (E, F) Simulated ultrasound, optimistic physics. (G) Training needed for the goal; gray, not reached with 15\,h; dots, participants; lines, medians. (H) Largest uniform noise (SNR change) at which the goal is still reached; $\blacktriangle$, beyond $-18$\,dB; $\blacktriangledown$, not reached.}
\label{fig:learning}
\end{figure*}

\section{Methods}
\label{sec:methods}

\subsection{Data and readouts}

In the public fMRI data of LeBel et al.\ \cite{lebel2023} (OpenNeuro ds003020), each of three participants (S1 to S3) listened to 82 training stories (15.25\,h) and to one 10-min test story, repeated ten times. Every readout saw the same voxels, chosen from anatomy alone before any story data were touched: auditory cortex, the LanA language network (probability $\geq 0.2$ \cite{lipkin2022}) and the speech-motor areas 6v, 6r, 55b, 44 and 45 \cite{glasser2016} of both hemispheres (9,694 to 12,113 voxels). The decoder of Tang et al.\ \cite{tang2023} (GPT-based encoding model, noise model, beam search) ran with its authors' settings (a beam of 200) and the most favorable word timing available (predicted from clean data with all 15\,h). The encode-then-compare verifier scores the likelihood of the recording under the response that the decoder's models predict for a candidate. The contrastive verifier projects the window's principal components linearly onto a sentence-embedding space and is trained with a contrastive loss. We fitted every readout anew on cloud GPUs for each condition and amount of training data (the opening $k$ of the 82 stories; 11 amounts from 11\,min for the verifiers, 9 from 49\,min for the decoder).

\subsection{Evidence in bits and goals}

Readouts were scored in bits per 20-s window of a held-out story. A verifier ranks the true segment among $P$ segments of other stories (timing-matched for encode-then-compare): evidence is $\log_2[(P+1)/\mathrm{rank}]$ minus the same with time-shifted brain data, so chance is 0 bits. As a control against speaker and session cues, the wrong options were also drawn from the same held-out story, none within 20\,s of the true window. For the decoder we counted, among 4,000 texts its language model wrote without brain data, those at least as close to the heard words (BERTScore \cite{zhang2020}) as the decoded text; evidence is $\log_2[4{,}001/(1+\mathrm{count})]$ minus its brain-free expectation. The goal is 1 bit for a verifier; for the decoder it is the evidence of the published decoder's median window with 15\,h of clean data (4.6 to 5.7 bits), the text quality it is built to deliver. We measured learning speed as the least-squares slope of evidence against $\log_{10}$ training time between 49\,min and 15\,h, in bits per tenfold, and compared readouts by a moving-block bootstrap over test windows. Noise is given as the change in each voxel's SNR relative to clean fMRI, in dB; $-6$\,dB means four times the noise power. Uniform noise with the fMRI noise's spatial structure changed every voxel's SNR by $-6$, $-12$ or $-18$\,dB in training and test data. Training needed and noise survived were interpolated.

\subsection{Transcranial noise model}

The simulation predicts each voxel's SNR for a power Doppler recording from a probe on the temple and adds Gaussian noise to that voxel's real fMRI time series until its SNR equals the prediction. It forms no ultrasound image. The left temporal window was drawn on each participant's scalp, from 25\,mm behind the ear canal to the temporal pole and up to 5\,cm above the zygomatic arch (32 to 36\,cm$^2$). In the two-probe case a second probe sat on the right temple, and each voxel took the nearer one.

The budget assumes 2\,MHz, a 20-mm aperture ($-2$\,dB) and skull aberration corrected to a 0.4-dB residual. The echo SNR of one Doppler sample per resolution cell is $1/s = 1/s_n + 1/s_c + 1/s_r$: receiver (thermal) noise $s_n$, residual tissue clutter $s_c$ (an assumed level, $\pm 10$\,dB) and reverberation $s_r$. Bone and tissue were modeled as attenuation only, with one-way coefficients doubled for the round trip: compact bone $2.1 f^{1.1}$\,dB/mm plus 2.9\,dB for its two surfaces \cite{webb2021}, brain $0.59 f^{1.3}$\,dB/cm \cite{itis2022}, scalp 1\,dB/cm/MHz. Our measure of signal is $d'$, the speech response divided by its noise in one 2-s scan (one fMRI volume, the repetition time TR) of a 27-mm$^3$ voxel ($d' = 1$: the response is as large as its noise). Following the physiological-noise model of fMRI \cite{kruger2001},
\begin{equation}
d' = \frac{\delta\sqrt{K}\,\mathrm{SNR}_0}{\sqrt{1+\lambda^2 K\,\mathrm{SNR}_0^2}}, \qquad \mathrm{SNR}_0 = \sqrt{N}\,\frac{s}{1+s},
\label{eq:dprime}
\end{equation}
where $\mathrm{SNR}_0$ is how precisely one 1-s power-Doppler image reads blood volume, $N$ its independent samples within one image, $K = 2$ images per scan, $\delta$ the fractional blood-volume response and $\lambda$ physiological noise. Two corrections turn $d'$ into what the readouts receive: only the part of the response that differs between sentences helps ($\rho = 0.26$, $-11.7$\,dB; from released human BOLD data we computed 0.07 to 0.65 \cite{normanhaignere2015,tuckute2024})\footnote{$\rho$: SD of the stimulus-specific response over the mean response. Human BOLD: 0.29 (auditory cortex, natural sounds), 0.07 (speech only) \cite{normanhaignere2015}, 0.65 (language network, sentences) \cite{tuckute2024}; monkey functional ultrasound: 0.26 \cite{griggs2024}, used here.}, and the fMRI voxel is smaller ($-1.8$\,dB). Together they give the SNR of one 2-s scan of an fMRI voxel, $\tilde d = \rho\,\sqrt{17.8/27}\;d' \approx 0.21\,d'$, 13.5\,dB below $d'$. A Monte Carlo of 20,000 draws, each input drawn independently and log-uniformly over its range (bone thickness uniformly), spans echo-SNR calibration ($\pm 10$\,dB), temporal bone thickness (2.0 to 3.8\,mm \cite{he2022}), $\delta$ (2 to 10\%), blood speed (1.5 to 6\,mm/s), clutter and reverberation leakage ($\pm 10$\,dB) and $\lambda$ (0.05 to 0.20); its 10th, 50th and 90th percentiles per distance define pessimistic, median and optimistic physics. They express uncertainty about the physics, not scatter between voxels. We kept voxels 1 to 8\,cm from the probe whose predicted SNR was above $-60$\,dB. For each, the ten repeats of the test story give its real fMRI SNR $r_v$ (the power of what repeats over the power of what does not)\footnote{The repeats only calibrate the added noise, offline; every readout was trained and scored on single trials.} and its noise variance $\sigma_v^2$; as ten repeats give noisy estimates, $r_v$ was shrunk toward the distribution over all voxels (empirical Bayes). Adding noise of variance $f_v\sigma_v^2$ lowers the SNR to $r_v/(1+f_v)$, so
\begin{equation}
f_v = \max\!\left(0,\; r_v/\tilde d_v^{\,2} - 1\right)
\label{eq:fv}
\end{equation}
brings every voxel to its prediction; voxels already below it receive none. The noise has the fMRI noise's spatial structure, is white in time and is independent for every story and repeat.

We did not simulate wave propagation. A link budget suits this question because per-voxel signal-to-noise through the temporal bone is dominated by attenuation; Vienneau et al.\ \cite{vienneau2026} report that aberration correction recovers only about 2 to 4\,dB, small compared with the attenuation loss. The Monte Carlo spans the uncertainty in these six inputs, which a full-wave simulation on one skull would not capture. The model has not been validated against task-evoked transcranial recordings, which do not yet exist; it is meant as a first step that motivates the field to make them.

The simulated recording keeps fMRI's 2-s scans, 2.6-mm voxels (17.8\,mm$^3$) and whole-cortex coverage. Some assumptions make it better than a real recording would be: window-quality bone under the whole probe, sound meeting the bone head-on, assumed clutter, continuous recording, and all reachable cortex instead of one imaging plane. Others make it worse: a single 2-MHz frequency and no coded pulses or contrast agent. No specific array, steering scheme or frame rate was modeled; the probe is idealized and samples every reachable voxel, so the result is an upper bound on one probe.

\textbf{Added gain.} To ask how much stronger the signal must be, we repeated the verifiers under median physics with every voxel's target $\tilde d_v$ raised by 6, 12 or 18\,dB (49\,min to 15\,h).

\section{Results}
\label{sec:results}

\subsection{What the temporal window delivers}

Through the temporal bone the echo from a voxel 5\,cm deep returns about 10\,dB below the receiver noise, mostly because of the bone, and one 2-s scan carries a speech signal 22 to 24\,dB below fMRI's (Fig.~\ref{fig:imaging}A, B). Across the readouts' voxels, one probe changes the SNR by $-11$ to $-14$\,dB under optimistic physics, close to the uniform $-12$\,dB condition, and by $-23$ to $-25$\,dB under median physics; under pessimistic physics only voxels 1.9 to 4.9\,cm away remain (Fig.~\ref{fig:imaging}C).

\subsection{Data needed, with and without uniform noise}

In clean data both verifiers passed 1 bit with the least training tried, or close to it (11 to 12\,min); the decoder reached its goal with 3.2, 6.2 and 6.6\,h (Fig.~\ref{fig:learning}A, G), 16 to over 30 times more. Every readout gained 2.6 to 4.2 bits per tenfold more data, so the verifiers' lead came from needing about 4 fewer bits (roughly twentyfold less data). On the 10,000 voxels its encoding model predicted best, the decoder learned as fast, gained about 0.75 bits at 15\,h and needed 3.1 to 5.5\,h, so the voxel set does not explain the gap.

With added noise the learning speeds diverged (Fig.~\ref{fig:learning}B to D). At $-6$\,dB the decoder gained 1.4 to 2.4 bits per tenfold, 1.0 to 1.8 less than in clean data and less than encode-then-compare in every participant (3.6 to 4.1); the verifiers lost at most 0.4. At $-12$\,dB the contrastive verifier gained 1.7 to 2.2, reached 1 bit with 0.9 to 2.1\,h and kept 2.7 to 3.6 bits with same-story wrong options (5.0 to 5.5 in clean data); encode-then-compare gained 1.2 to 2.5 and needed 3.7 to 6.3\,h, and the decoder gained 0.1 to 1.3 and scored 0.8 to 2.3 bits at 15\,h, slower than the contrastive verifier in all three participants and than encode-then-compare in two. At $-18$\,dB the contrastive verifier still gained 0.6 to 1.2 and reached its goal in two participants (3.4 and 6.9\,h; 0.87 bits in the third), while encode-then-compare stalled (0.13 to 0.17 bits per tenfold) and the decoder gained 0.0 to 0.5 and scored 0.4 to 0.9 bits at 15\,h. From $-6$\,dB on, the decoder never reached its goal (Fig.~\ref{fig:learning}G). With 15\,h the decoder kept its goal only to $-0.8$ to $-2.3$\,dB, encode-then-compare to $-14.5$ to $-16.5$\,dB and the contrastive verifier to $-17.6$\,dB or beyond (Fig.~\ref{fig:learning}H): in every participant the verifiers withstood 13 to over 17\,dB more noise than the decoder.

\subsection{Simulated ultrasound}

With simulated ultrasound and optimistic physics both verifiers reached their goal: the contrastive verifier after 2.2 to 2.3\,h of training with one probe and 1.7 to 1.8\,h with two, encode-then-compare after 5.6 to 10.2\,h and 6.3 to 7.8\,h (Fig.~\ref{fig:learning}E to G). They gained 1.4 to 1.8 and 0.7 to 1.4 bits per tenfold, steeper than the decoder's in all six cases (three participants, one or two probes). The decoder learned little (0.3 to 1.0 bits per tenfold): in three of the six cases, all with one probe, the confidence interval of its slope included zero, and it scored 0.5 to 1.4 bits at 15\,h.

\subsection{How much stronger the signal must be}

\begin{figure}[t]
\centering
\includegraphics[width=\columnwidth]{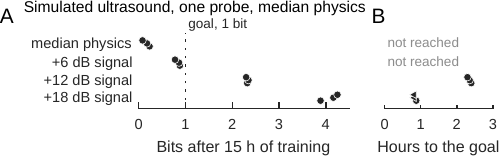}
\caption{\textbf{Under simulated median physics, verification needs about 12\,dB more SNR.} Contrastive verifier; the lower rows raise every voxel's SNR by the stated amount; dots, participants. (A) Evidence after 15\,h of training; dotted, the 1-bit goal. (B) Training needed for the goal; triangles, reached with the least tried (0.8\,h).}
\label{fig:e3}
\end{figure}

Under simulated median physics neither verifier came near its goal (at most 0.23 bits). Raising every voxel's functional SNR by 6\,dB fell just short (0.8 to 0.9 bits); raising it by 12\,dB was enough: the contrastive verifier then reached 1 bit with 2.3 to 2.4\,h of training, and with 18\,dB in 0.9\,h or less (Fig.~\ref{fig:e3}). This is comparable to the 11 to 18\,dB that coded excitation has added through the temporal window \cite{vienneau2023}.

\section{Discussion and limitations}
\label{sec:discussion}

Because the slopes themselves diverged, the verifiers' advantage under noise holds wherever the goals are placed. These results are consistent with the hypothesis that the evidence needed to confirm a sentence may lie in coarse, redundant components of the cortical response that survive noise, whereas generating text depends on fine components that noise removes first.

As a rough check, scaling the published transcranial breath-hold signal \cite{vienneau2026} to a sentence-sized response gives about $-40$ to $-31$\,dB per 2-s sample, between our median ($-41.4$\,dB) and optimistic (about $-30$\,dB) cases: an order-of-magnitude consistency, not a validation (a global response in deep arteries, recorded with coded excitation and reported in log-compressed units), good to about $\pm 10$\,dB.

Coded pulses could supply the missing 12\,dB, but the 65-chip code of \cite{vienneau2026} cannot see its first 2\,cm, which hold 11 to 23\% of the readouts' voxels. The data come from three people, and fMRI is a proxy for blood volume. Verification was tested only at the level of gist, against other stories or other windows of the same story, rather than paraphrases or negations, and the decoder was given the best word timing. Next are a full-wave simulation on realistic skull models \cite{treeby2010,pichardo2023,iacono2015} and a direct transcranial measurement with a probe on each temporal window.

\clearpage

\section{Compliance with ethical standards}
\label{sec:ethics}

This study was conducted retrospectively using human subject data made available in open access by LeBel et al.\ (OpenNeuro ds003020), collected under the approval of the original institution's IRB. Ethical approval was not required for this secondary analysis of de-identified open data.

\section{Acknowledgments}
\label{sec:acknowledgments}

This work was done independently, without institutional resources or funding. Conflict of interest: none. AI use: Claude (Anthropic) was used for coding and figure generation; all content and numbers were checked by the author.

\noindent{\raggedright Code: \href{https://github.com/deShal3v/verify-vs-decode-ultrasound}{\nolinkurl{github.com/deShal3v/verify-vs-decode-ultrasound}} (\mbox{v1.0-isbi}, doi:\href{https://doi.org/10.5281/zenodo.23527277}{10.5281/zenodo.23527277}).\par}

\bibliographystyle{IEEEbib}
\bibliography{refs}

\end{document}